\documentstyle[12pt,epsf]{article}    
\begin{document}
\begin{center} {\bf  High-Temperature Effective Potential of Noncommutative Scalar
Field Theory:\\ Reduction of  Degrees of Freedom by Noncommutativity\\ }
                                                  
\vspace{1cm}

                      Wung-Hong Huang\\
                       Department of Physics\\
                       National Cheng Kung University\\
                       Tainan,70101,Taiwan\\

\end{center}
\vspace{1cm}

    The renormalization of effective potentials for the noncommutative scalar field
theory at high temperature are investigated to the two-loop approximation.  The
Feynman diagrams in evaluating the effective potential may be classified into two
types: the planar diagrams and nonplanar diagrams.   The nonplanar diagrams, which
depend on the parameter of  noncommutativity, do not appear in the one-loop
potential.   Despite their appearance in the two-loop level, they do not have an
inclination to restore the symmetry breaking in the tree level, in contrast to the
planar diagrams.   This phenomenon is explained as a consequence of the drastic
reduction of the degrees of freedom in the nonplanar diagrams when the thermal
wavelength is smaller than the noncommutativity scale.    Our results show that the
nonplanar two-loop contribution to the effective potential can be neglected  in
comparsion with that from the planar diagrams.

\vspace{1cm}
\begin{flushleft}
E-mail:  whhwung@mail.ncku.edu.tw\\
Keywords: Superstrings and Heterotic String, Non-commutative Geometry.
 PACS:11.10.Lm; 11.10.Kk; 11.10.Wx; 12.38.Bx
\end{flushleft}


\newpage
\section  {Introduction}

   Field theory on noncommutative space (or spacetime)  has received a great deal of
attention recently.     Initially, Connes, Douglas and Schwarz [1] showed that the
supersymmetric gauge theory on noncommutative torus is naturally related to the
compactification of Matrix theory.   From string theory, it was also found that the
end points of the open strings trapped on a D-brane with a nonzero NSNS two form
B-field background turns out to be noncommuting [2-4].  As it has proved to arise
naturally in the string/M theory the noncommutative field theory has attracted many
researchers [5-18].    

 The quantum field theories on the noncommutative space (or spacetime) have been
pursued via perturbative analysis over diverse model.   The noncommutative scalar
field theory have been considered in [5-8].  It has been shown that this theory is
renormalizable up to two loops.  The pure noncommutative gauge theory has also been
shown to be renormalizable up to one loop [9,10].   Historically it was a hoped that
introducing a minimum scale to deform the geometry in the small spacetime would be
possible to cure the quantum-field divergences, especially in the gravity theory
[11,12].   Although the noncommutative field theory turns out to exhibit the same
divergence as the commutative one [5], it is of interests in its own right.  The
special properties of the nonlocality (they contain infinite order derivatives) and
the existence of  a new parameter (the noncommutativity parameter, $\theta$) in the
noncommutative field theories make them challenging and lead to some fascinating
behaviors.

   A distinct characteristic of the  noncommutative field theories, found by
Minwalla, Raamsdonk and Seiberg [6], is the mixing of ultraviolet (UV) and infrared
(IR) divergences reminiscent of the UV/IR connection of the string theory.   The
nontrivial mixing between UV and IR is very special and may be explained as the
nonlocality shown in the noncommutative spacetime [6]. 

  In recent papers [13,14], it has also been found that the noncommutativity in the
extra spaces may be used to stabilize the radius of extra space by the Casimir
effect.   It seems that the dimensional parameter of space noncommutativity can
provide a minimum scale to protect the collapse of the extra space in some systems. 

Another interesting characteric of the appearance of a minimum scale is also found
in the finite-temperature noncommutative field theories [15-17].   Fischler {\it el.
al.} [15] showed that, at high temperature for which the thermal wavelength is small
than the noncommutativity scale, there is no way to distinguish and count the
contributions of modes to the free energy.   Thus, there is a drastic reduction of
the degrees of freedom in the non-planar contribution to the thermodynamical
potential at high temperature.

    In this paper we will investigate the renormalized two-loop effective potential 
for noncommutative scalar field theory in the high-temperature limit.    We will see
that the property of the reduction of the degrees of freedom found in [15] can also
be seen in the effective potential.  Note that the free energy evaluated in [15] is
given to the second order of coupling constant $\lambda$, while the effective
potential evaluated in this paper is given to the second order of  $\hbar$ and to
all order in the coupling  $\lambda$.   As the space-time noncommutativity
($\theta_{0i} \ne 0$) will lead to infinite number of time derivative, it will
render the field theory nonlocal in time and the causality may be violated at the
quantum level [18].  Therefore we shall in this paper consider only space
noncommutative theory. 

  In section 2, we briefly review the relation between the field theory on
noncommutative spacetime and the noncommutative field theory.   The path-integration
formulation of Jackiw [19,20], which is used to evaluate the effective potential, is
then extend to the noncommutative theory and the Feynman diagrams ae derived.   In
this section we see that the spacetime noncommutativity does not affect the one-loop
potential.   Thus the radiatively symmetry breaking is blind to the noncommutativity
at this level [5,8].   It is seen that this property is independent of the spacetime
dimension and irrelevant to temperature.  

    The Feynman rule derived in section 2 is then used to analyze the two-loop
diagram of the $\lambda \phi^4$ theory at high temperature.  The Feynman diagrams
therein may be classified as two types:  the planar diagrams and nonplanar diagrams.
  The nonplanar diagrams are the parts which will depend on the parameter of space
noncommutativity.   They appear in the two-loop level but do not have an inclination
to restore the symmetry breaking in the tree level.  This property is in contrast to
the conventional believing that high temperature could restore the symmetry
breaking.  To explain this phenomenon we compare the result with the effective
potential in the zero-space dimension and explain this property as a consequence of
drastic reduction of the degrees of freedom in the nonplanar diagrams at high
temperature in which the thermal wavelength is smaller than the noncommutativity
scale [15].   Our results, however,  show that the nonplanar two-loop contribution
to the effective potential can be neglected in comparing to that from the planar
diagrams.

  To confirm the above argument of the drastic reduction of the degrees of freedom
in the nonplanar diagram at high temperature, we present in section 4 an analysis of
the two-loop effective potential of the $\lambda \phi^3$ theory at high temperature.
  In the last section we give a short conclusion. 


\section  {Formulation }
\subsection {Quantum Field on Noncommutative Spacetime}

   We consider the noncommutative geometry ${\cal R}^{n}$ defined in $n$ dimensions
with the commutations
$$  \left[ {\hat x}^{\mu},{\hat x}^{\nu} \right] = i \theta^{\mu\nu} ,
\eqno{(2.1)}$$
\noindent
where $\theta^{\mu\nu}$are real C-numbers.  Given this algebra  we can follow the
method of Weyl [21] to describe the functions living on the noncommutative
spacetime.    The method is to define the function 
$${\hat f}(\hat x) = {1\over (2 \pi)^{n/2}} \int d^{n} k \hspace{5pt} e^{i k_{\mu}
{\hat x}^{\mu}} {\tilde f}(k),   \eqno{(2.2)}$$
where ${\tilde f}(k)$ is the Fourier transform of $f(x)$ :
$${\tilde f}(k) = {1\over (2 \pi)^{n/2}} \int d^{n} x ~ e^{-i k_{\mu} x^{\mu}}~ f(x)
,  \eqno{(2.3)}$$
in which $x$ is the commuting variable corresponding the noncommuting variable
${\hat x}$.  This definition uniquely associates a function ${\hat f}(\hat x)$
living on the the noncommutative spacetime with a function $f( x)$ living on the
commutative spacetime.    From the above definition the product of two functions
${\hat f}(\hat x)$ and ${\hat g}(\hat x)$ then becomes

$$ {\hat f}(\hat x) \cdot {\hat g}(\hat x) ={1\over (2 \pi)^{n}} \int d^{n}k d^{n}p
~ e^{i k_{\mu} {\hat x}^{\mu}} ~ e^{i p_{\nu} {\hat x}^{\nu}} ~ {\tilde f}(k) ~
{\tilde g}(p)  \hspace{3cm}$$ 
$$ = {1\over (2 \pi)^{n}} \int d^{n}k d^{n}p ~ e^{i (k_{\mu} + 
p_{\mu}) {\hat x}^{\mu} - {i\over2} k_{\mu} \theta^{\mu\nu} p_{\nu}} ~ \tilde{f}(k)
\tilde{g}(p). \eqno{(2.4)}$$
\\
To obtain the above relation we have used the  Baker-Campbell-Hausdorff formula
$$e^{A} e^{B} = exp\left ( A+B +{1\over2} [A,B] + +{1\over12} [A,[A,B]]+{1\over12}
[B,[B,A]] + \ldots\right ) ,$$
and the fact that the commutators $\theta^{\mu\nu}$are constants, thus the higher
commutators vanish.  From Eq.(2.4), we see that once we define the Moyal product
($\ast$) [22]

$$f(x) \ast g(x) = {1\over (2 \pi)^{n}} \int d^{n}k d^{n}p ~ e^{i 
(k_{\mu} + p_{\mu}) x^{\mu}} e^{- {i\over2} k_{\mu} \theta^{\mu\nu} p_{\nu}} ~ 
\tilde{f}(k) \tilde{g}(p) $$

$$ = e^{{i\over 2} \theta^{\mu\nu} {\partial\over \partial y^{\mu}} 
{\partial\over \partial z^{\nu}} } f(y) g(z) |_{y,z\rightarrow x} , \hspace{1.5cm}
\eqno{(2.5)}$$  \\  
\noindent
we can establish a homomorphism, ${\hat f}(\hat x) \cdot {\hat g}(\hat x)  =
f(x)\ast g(x)$. This homomorphism allows us to view the algebra of functions on
noncommutative spacetime ${\cal R}^{n}$ as the algebra of the ordinary functions on
commutative ${\cal R}^{n}$ with the Moyal $\ast$-product instead with the usual
pointwise product. 

    Therefore, in investigating the field theory on noncommutative spacetime  we can
alway work on a usual commutative spacetime in which the multiplication operator is
replaced by the so called Moyal $\ast$ product; in other words,  we are going to
study the problem of noncommutative field theory.

   Note that the  Moyal $\ast$ product satisfies the law of associativity:
  $$\left ( f(x) \ast g(x) \right ) \ast h(x) = f(x) \ast \left ( g(x) \ast h(x)
\right ). \eqno{(2.6a)}$$
Under the integral it also has a property:
 $$\int d^n x \left ( f(x) \ast g(x) \right ) = \int d^n x \left ( f(x) \cdot g(x)
\right ), \eqno{(2.6b)} $$
as the noncommutativity $\theta $ is an antisymmetric matrix.


\subsection{Path-Integral Formulation of Effective Potential}

   In this section we will consider the $\lambda \phi^4$ theory on a noncommutative
spacetime.   Using the above prescription we can write the Lagrangian as 

$$S[\phi] = \int d^n x ~{\cal L}(\phi) = \int d^n x ~ \left [~{1\over 2}
\partial_\mu \phi ~  \partial^ \mu \phi - {1\over 2} m^2 \phi ^2 - {\lambda\over
4!}\phi \ast \phi \ast \phi
\ast  \phi \right] .  \eqno{(2.7)}$$
\noindent
We will evaluate the renormalized effective potential of the above model along the
path-integration formulation of Jackiw [19,20].

   First, we assume that there exists a stationary point at which $\phi $ is a
constant field $\phi_0$.   Thus 
$$ \left.{\delta S \over \delta \phi} \right | _ {\phi_0} = 0.$$
Next, we expand the Lagrangian about the stationary point the action becomes

$$S[\phi] = S[\phi_0] + {1\over 2}\int d^n x ~ d^n y ~\tilde{\phi}(x) ~ \star 
\tilde{\phi}(y) ~ \left .{\delta^2 S \over {\delta \phi (x) \delta \phi
(y)}}  \right | _{\phi_0}+ \int d^n x ~\tilde {\cal L_I} (\tilde \phi ,\phi
_0 ), \eqno{(2.8)}  $$
\
in which $\tilde{\phi} \equiv \phi - \phi_0 $ and $\tilde {\cal L_I}
(\tilde \phi ,\phi _0 )$ can be found from  the Lagrangian Eq.(2.7).   
Then we use the propagator defined by  

$$i D^{-1}(\phi_0;p) = \int d^n p e^{i p x}~ i D^{-1}(\phi_0;x,0),
\eqno{(2.9a)}\\$$
$$ i D^{-1}(\phi_0;x,y) = \left .{\delta^2 S \over {\delta \phi (x) \delta
\phi (y)}}  \right | _{\phi_0}, ~~~~~~~~ \eqno{(2.9b)}  $$
the effective potential $V(\phi_0)$ is found to be [19]

$$V(\phi_0) = V_0 (\phi_0) - {1\over 2} i ~\hbar \int {d^n p \over {(2
\pi)^n}} \ln det \left[ i D^{-1}(\phi_0;p)\right] + i ~ \hbar <exp \left ({i \over
\hbar
}\int d^n x ~  \tilde {\cal L_I} (\tilde \phi ,\phi _0 ) \right )>. 
\eqno{(2.10)} $$
\\
For the theory at finite temperature $T = 1/\beta$ we shall take the following
substitutions [20]:

 $$p_0 \rightarrow {2 \pi p_0\over \beta },  \eqno{(2.11)}$$
 $$\int d^n p \rightarrow  {2 \pi \over \beta} \sum_{p_0} \int d^{n-1} {\bf p}
\eqno{(2.12)},$$
in which $p_0$ is an integral.

The first term in Eq.(2.10) is the classical potential which can be read from
Eq.(2.7).   The second term is the one-loop contribution which comes from the second
term in Eq.(2.8).  The elementary property (2.6b) implies that the Moyal
$\ast$-product in the second term of Eq.(2.8) can be dropped.    Thus we see an
interesting property that the noncommutativity of spacetime dose not affect the
potential in the one-loop level.    This property has been found by Campbell and
Kaminsky [5] in the investigation of the tadpole diagram in the linear sigma model. 
 It is easy to see that this property is independent of the spacetime dimension and
irrelevant to the temperature.  

    The third term in Eq.(2.10) is the higher-loop contribution of the effective
potential.   To obtain it one shall evaluate the expectation value of the third term
in Eq.(2.8) by the Feynman rule, with $D(\phi_0;p)$, defined in Eq.(2.9),  as the
propagator and keep only the connected single-particle irreducible graphs [19,20]. 

 The Feynman rules including the propagator and vertices are shown in figure 1.
\\
\unitlength 2mm
\begin {picture}(30,5)
\put(15,0){\vector(1,0){7}}
\put(20,0){\line(1,0){8}}
\put(20,2){$ p_0, {\bf p}$}
\end {picture}
\hspace{1mm} $= {1\over {({2\pi p_0\over\beta})^2 + {\bf p}^2} +M^2} $
\\
\begin {picture}(50,12)
\put(15,0){\line(1,1){10}}
\put(15,10){\line(1,-1){10}}
\put(15,0){\vector(1,1){3}}
\put(15,10){\vector(1,-1){3}}
\put(25,0){\vector(-1,1){3}}
\put(25,10){\vector(-1,-1){3}}
\put(17,0){$p_1$}
\put(17,10){$p_4$}
\put(22,0){$p_2$}
\put(22,10){$p_3$}
\put(30,5){$ = -  6 \lambda V({\bf p}_1,{\bf p}_2,{\bf p}_3,{\bf p}_4) $}
\end {picture}
\\

\begin {picture}(50,12)
\put(15,0){\line(1,1){5}}
\put(20,5){\line(1,-1){5}}
\put(20,5){\line(0,1){5}}
\put(15,0){\vector(1,1){3}}
\put(25,0){\vector(-1,1){3}}
\put(20,10){\vector(0,-1){3}}
\put(17,0){$p_1$}
\put(22,0){$p_2$}
\put(21,8){$p_3$}
\put(30,5){$=  -  6 \phi_0 \lambda V({\bf p}_1,{\bf p}_2,{\bf p}_3) $}
\end{picture}
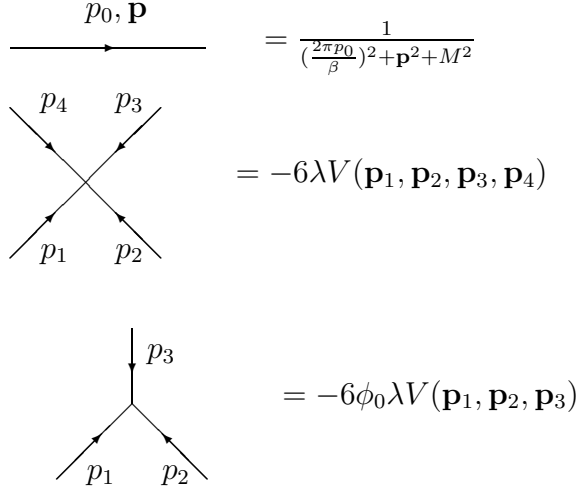
\begin{center} {Figure 1.  Feynman rules : propagator and vertices}\end{center}
\noindent
In figure 1 we define  
 $$M^2 =m^2 + {1 \over 2} \lambda \phi _0 ^2 ~,  \eqno{(2.13)}$$
and 
 $$V({\bf p}_a,{\bf p}_b, ...) = exp\left [{-i\over2}{\displaystyle
\sum_{a<b}} ({\bf p}_a)_ i \theta^{ij} ({\bf p}_b)_j \right] .
\eqno{(2.14)}$$ 
\\
Note that we consider only space noncommutative theories (i.e. $\theta_{0i}
= 0$) for the unitarity requirement [18].   

  In the following sections we will use the above Feynman rules to
evaluated  the two-loop diagram for the system at high
temperature.    We will use the zeta-function regularization method [23] to
perform the summations over the integral values of
$p_0$ and $k_0$.


\section {Two-Loop Corrections: $\lambda \phi^4$ Theory}

    From the Feynman rule we see that the diagrams in the two-loop level 
can be divided into two types: planar diagrams and nonplanar diagrams.   The
contributions of the effective potentials from the planar diagrams come from the two
diagrams [19,20,8] shown in figure 2.

\begin {picture}(15,8)
\put(11,2){\circle{4}}
\put(9,2){\line (1,0 ){4}}
\put (17,2){= $I_2^P$}
\end {picture}
\begin {picture}(29,8)
\put(19,2){\circle{4}}
\put(23.2,2){\circle{4}}
\put (27,2){= $I_1^P$}
\end {picture}
\begin{center}
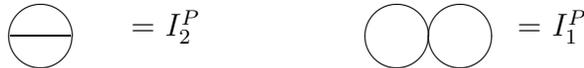
{Figrue 2. Planar two-loop contributions to the effective potential}
\end{center}
\noindent
For the theory in 1+3 dimensions the effective potentials evaluated from figure 2
are
 
$$I^P_1 = ~{2\over 3}~{{\hbar ^2} \over 24} {\lambda \over\beta^2}~ 
{\displaystyle \sum_{k_0 }} \int  {d^3 {\bf k} \over
{(2\pi)^3}}{1\over({2\pi k_0\over\beta })^2 + {\bf k}^2 +M^2} 
{\displaystyle \sum_{p_0 }} \int  {d^3 {\bf p} \over
{(2\pi)^3}}{1\over({2\pi p_0\over\beta})^2 + {\bf p}^2 +M^2} ,   
\eqno{(3.1)}$$

  $$ I^P_2= - ~ {1\over 2} ~ {{\hbar ^2} \over 36} {\lambda ^2
\phi_0^2 \over \beta^2}{\displaystyle \sum_{k_0 }}{\displaystyle \sum_{p_0
}} \int {d^3 {\bf k} \over {(2\pi)^3}}\int  {d^3 {\bf p} \over {(2\pi)^3}}
{1\over({2\pi k_0\over \beta})^2 + {\bf k}^2 +M^2} \times \hspace{3cm}$$
$$  {1\over({2\pi p_0\over \beta})^2 + {\bf p}^2 +M^2} {1\over({2\pi
k_0\over \beta}+ {2\pi p_0\over \beta})^2 + ({\bf p +\bf k})^2
+M^2}. \eqno{(3.2)}$$
\\
The contributions from the nonplanar diagrams are like those in the planar diagrams,
but with an extra factor $e^{i {\bf k}_i\theta^{ij} {\bf p}_j}$.   They read

$$I_1^N = ~ {1\over 3} ~ {{\hbar ^2} \over 24} {\lambda \over\beta^2}~ 
{\displaystyle \sum_{k_0 }}{\displaystyle \sum_{p_0 }} \int  {d^3 {\bf k} \over
{(2\pi)^3}}{d^3 {\bf p} \over {(2\pi)^3}}{e^{i {\bf k}_i\theta^{ij} {\bf
p}_j}\over{[({2\pi k_0\over\beta })^2 + {\bf k}^2 +M^2}]  [({2\pi p_0\over\beta})^2
+ {\bf p}^2 +M^2]} ,    \eqno{(3.3)}$$
\\
  $$I_2^N = - ~ {1\over 2} ~ {{\hbar ^2} \over 36}  {\lambda ^2
\phi_0^2 \over \beta^2}{\displaystyle \sum_{k_0 }}{\displaystyle \sum_{p_0 }}
\int {d^3 {\bf k} \over {(2\pi)^3}} {d^3 {\bf p} \over {(2\pi)^3}} {1\over({2\pi
k_0\over \beta})^2 + {\bf k}^2 +M^2} \times \hspace{3.5cm}$$

$$  {1\over({2\pi p_0\over \beta})^2 + {\bf p}^2 +M^2} {e^{i {\bf k}_i\theta^{ij}
{\bf p}_j}\over({2\pi k_0\over \beta}+ {2\pi p_0\over \beta})^2 + ({\bf p +\bf k})^2
+M^2}.\eqno{(3.4)}$$
\\
Note that the factors ${2\over 3}$ (${1\over 3}$) appearing in Eqs. (3.1)
((3.3)) means that the associated planar (nonplanar) diagram will be with
2/3 (1/3) weight of the commutative graph.   And the factor ${1\over 2}$
appearing in Eqs. (3.2) and (3.4) means that the associated planar and
nonplanar diagrams will both have weight of 1/2 of the commutative graph.  
The counting rule has been detailed by Campbell and Kaminsky [5] in
the investigation of the linear sigma model.   Let us describe it again for
completeness.  

  The diagram (3.1) has a single vertex, and so has a phase factor
$V(p,k,-k,-p)$.   Of the six possible orderings (modulo cyclic permutation)
of the set $\{p,k,-k,-p\}$, four have a trivial phase factor, and two
have a phase of either $e^{i {\bf k}_i \theta^{ij} {\bf p}_j}$ or $e^{-  i {\bf k}_i
\theta^{ij} {\bf p}_j}$ (which are the same under the integral over the loop momenta
$k$).  Thus the planar diagrams will have 4/6=2/3 weight and the nonplanar diagrams
will have 2/6=1/3 weight with respect to the commutative graph.  .  

   The diagram (3.2) has two vertices, and we pick up the phase factor
$V(p,k,-p-k)  V(-k,-p, p+k) $.   Each vertex has two orderings (modulo
cyclic permutation), for four combinations in total. Explicity evaluation leads to 
two having a trivial phase factor, and another two having a phase $e^{i {\bf k}_i
\theta^{ij} {\bf p}_j}$.   Thus the planar diagrams will with 2/4=1/2
weight and the nonplanar diagrams will with 2/4=1/2 weight with respect to
the commutative graph.  

    
\subsection {Nonplanar Diagrams}

We first analyze the nonplanar diagrams.   Using the Schwinger parameters,
$\alpha_1$ and $\alpha_2$, that in Eq. (3.3) can be expressed as 

$$ {1\over\beta^2}{\displaystyle \sum_{k_0 }}{\displaystyle \sum_{p_0 }} \int  d^3
{\bf k} ~ d^3 {\bf p}~ {e^{i {\bf k}_i\theta^{ij} {\bf p}_j}\over{[({2\pi
k_0\over\beta })^2 + {\bf k}^2 +M^2}]  [({2\pi p_0\over\beta})^2 + {\bf p}^2 +M^2]}
\hspace{5cm}$$
$$ = {1\over\beta^2}\int_0^\infty d\alpha_1 \int_0^\infty d \alpha_2 ~
{\displaystyle \sum_{p_0 }}{\displaystyle \sum_{l_0 }} ~ \int d^3 {\bf p}  d^3 {\bf
l}~ e^{- {1\over {4 \alpha_1}} \tilde {\bf p}_i \tilde {\bf p}^j} e^{-\alpha_1
[({2\pi p_0\over\beta })^2 + {\bf p}^2 +M^2]} ~ e^{-\alpha_2[({2\pi
l_0\over\beta})^2 + {\bf l}^2 +M^2]}$$
$$ = {1\over\beta^2}\int_0^\infty d\alpha_1 \int_0^\infty d \alpha_2 ~ ({\pi \over
\alpha_1})^{3\over 2} ~ e^{-(\alpha_1+\alpha_2) M^2}~\int d^3 {\bf p} ~ e^{- {1
\over {4\alpha_1}} \tilde {\bf p}_i \tilde {\bf p}^i} e^{-\alpha_2 {\bf p}^2 } ~
{\displaystyle \sum_{p_0 }} e^{-\alpha_1 ({2\pi p_0\over\beta})^2}  ~{\displaystyle
\sum_{l_0 }} e^{-\alpha_1 ({2\pi l_0\over\beta })^2},       
\eqno{(3.5)}$$
\\ 
in which $\tilde {\bf p}^i = \theta^{ij}{\bf p}_j$ and ${\bf l}_j = {\bf k}_j - {i
\over {2\alpha_1}}\tilde {\bf p}_j $.   

   In the same way, using the Feynman parameter $w$, the Schwinger parameters,
$\alpha_1$ and $\alpha_2$, that in Eq. (3.4) can be expressed as 

$$ {1\over\beta^2}{\displaystyle \sum_{k_0 }}{\displaystyle \sum_{p_0 }}
\int d^3 {\bf k} d^3 {\bf p} {1\over({2\pi k_0\over \beta})^2 + {\bf k}^2 +M^2}
{1\over({2\pi p_0\over \beta})^2 + {\bf p}^2 +M^2} {e^{i {\bf k}_i\theta^{ij} {\bf
p}_j}\over({2\pi k_0\over \beta}+ {2\pi p_0\over \beta})^2 + ({\bf p +\bf k})^2
+M^2} $$
$$ ={1\over\beta^2} \int _0^1 dw \int_0^\infty d\alpha_1 ~ \alpha_1 ~ \int_0^\infty
d \alpha_2 ~\int  d^3 {\bf l} ~ d^3 {\bf p} ~e^{- {1 \over {4\alpha_1}} \tilde {\bf
p}_i \tilde {\bf p}^i} ~ e^{- (\alpha_2+\alpha_1(w-w^2)){\bf p}^2} ~e^{- \alpha_1
{\bf l}^2} e^{-(\alpha_1+\alpha_2)M^2}\times$$
$$ {\displaystyle \sum_{k_0 }}{\displaystyle \sum_{p_0 }} ~ e^{-[\alpha_2 +
\alpha_1(w-w^2)]({2\pi p_0\over \beta })^2} ~ e^{-\alpha_1[{2\pi k_0\over \beta
}+(1-w){2\pi p_0\over \beta }]^2}$$
$$= {1\over\beta^2}\int _0^1 dw \int_0^\infty d\alpha_1 ~ \alpha_1~ \left({\pi\over
\alpha_1}\right)^{3\over 2}~ \int_0^\infty d \alpha_2 
~~e^{-(\alpha_1+\alpha_2)M^2}~\int d^3 {\bf p} ~e^{- {1 \over {4\alpha_1}} \tilde
{\bf p}_i \tilde {\bf p}^i} ~ e^{- (\alpha_2+\alpha_1(w-w^2)){\bf p}^2}\times$$
$$  {\displaystyle \sum_{k_0 }}{\displaystyle \sum_{p_0 }} ~ e^{-[\alpha_2 +
\alpha_1(w-w^2)]({2\pi p_0\over \beta })^2} ~ e^{-\alpha_1[{2\pi k_0\over \beta
}+(1-w){2\pi p_0\over \beta }]^2} ,\eqno{(3.6)}$$
\\ 
in which $\tilde {\bf p}^i = \theta^{ij}{\bf p}_j$ and ${\bf l}_i = {\bf k}_i -
(1-w){\bf p}_i - {i \over {2\alpha_1}}\tilde {\bf p}_i $. 

   We shall now integrate the momentum ${\bf p}$  in Eqs.(3.5) and (3.6). To do this
we first see that, because $\theta_{ij}$ is an antisymmetric matric the  $U_{ij}$
matric in the expression $\tilde {\bf p}_i \tilde {\bf p}^ i =  {\bf p}^i
\theta_{ij} \theta^ {jk} {\bf p}_k = {\bf p}^i U_i^k {\bf p}_k$  is  symmetric.  
Then because any real, symmetric matrix can be diagonalized by an orthogonal matrix
we can thus change the orthonormal variables ${\bf p}_i$ into another orthonormal
variables ${\bf h}_i$ and find that 

$${1 \over {4\alpha_1}} \tilde {\bf p}_i \tilde {\bf p}^i + \alpha_2 {\bf p}^2 =
\alpha_2 {\bf h}_1^2 + \left( \alpha_2 +
{\theta_{12}^2+\theta_{23}^2+\theta_{31}^2\over 4 \alpha_1}\right) ({\bf h}_2^2 +
{\bf h}_3^2) ,    \eqno{(3.7)} $$
$${1 \over {4\alpha_1}} \tilde {\bf p}_i \tilde {\bf p}^i + \left(\alpha_2
+\alpha_1(w-w^2)\right){\bf p}^2 = \left(\alpha_2 +\alpha_1(w-w^2)\right) {\bf
h}_1^2 +$$
$$ \hspace{4cm}\left( \left(\alpha_2 +\alpha_1(w-w^2)\right) + 
{\theta_{12}^2+\theta_{23}^2+\theta_{31}^2\over 4 \alpha_1}\right) ({\bf h}_2^2 +
{\bf h}_3^2).    \eqno{(3.8)} $$
\\
Note that the orthonormal variables ${\bf h}_i $ in Eq.(3.7) are different from
those in Eq.(3.8).

   Using the relations (3.7) and (3.8 ), Eqs.(3.5) and (3.6) become 

$$ {1\over\beta^2}\int_0^\infty d\alpha_1 \int_0^\infty d \alpha_2 ~ {\pi \over
\sqrt{\alpha_1 \alpha_2}} ~ {4 \pi^2\over 4 \alpha_1 \alpha_2+
(\theta_{12}^2+\theta_{23}^2+\theta_{31}^2)} ~e^{-(\alpha_1+\alpha_2) M^2} ~
{\displaystyle \sum_{p_0 }} e^{-\alpha_1 ({2\pi p_0\over\beta})^2}  ~{\displaystyle
\sum_{l_0 }} e^{-\alpha_1 ({2\pi l_0\over\beta })^2},       
\eqno{(3.9)}$$

$$ {1\over\beta^2}\int _0^1 dw \int_0^\infty d\alpha_1 \int_0^\infty d\alpha_2  {\pi
\sqrt\alpha_1\over\sqrt{\alpha_2 +\alpha_1(w-w^2)}} ~ {4 \pi^2
e^{-(\alpha_1+\alpha_2) M^2}\over (\theta_{12}^2+\theta_{23}^2+\theta_{31}^2) + 4
\alpha_1 (\alpha_2+\alpha_1(w-w^2)) }  \times$$
$$  {\displaystyle \sum_{k_0 }}{\displaystyle \sum_{p_0 }} ~ e^{-[\alpha_2 +
\alpha_1(w-w^2)]({2\pi p_0\over \beta })^2} ~ e^{-\alpha_1[{2\pi k_0\over \beta
}+(1-w){2\pi p_0\over \beta }]^2} .\eqno{(3.10)}$$
\\
respectively, after the integration of ${\bf h}_i$.

   To proceed, let us first investigate Eq.(3.9).   

$$ (3.9) ={1\over\beta^2}\int_0^\infty d\alpha_1 \int_0^\infty d \alpha_2 ~ {\pi
\over \sqrt{\alpha_1 \alpha_2}} ~ {4 \pi^2\over  \theta_{12}^2 + \theta_{23}^2 +
\theta_{31}^2}  \left [ 1 + \sum_{n=1} \left( {- 4 \alpha_2 \alpha_2 \over 
\theta_{12}^2 + \theta_{23}^2 + \theta_{31}^2} \right ) ^n   \right ] \times
\hspace{1cm} $$
$$~e^{-(\alpha_1+\alpha_2) M^2} ~ {\displaystyle \sum_{p_0 }} e^{-\alpha_1 ({2\pi
p_0\over\beta})^2}  ~{\displaystyle \sum_{l_0 }} e^{-\alpha_1 ({2\pi l_0\over\beta
})^2}$$
$$  = {4 \pi^4\over  \theta_{12}^2 + \theta_{23}^2 + \theta_{31}^2} ~ {1\over M^2
\beta^2} \left[ 1 + O\left( {\beta^2 \over (\theta_{12}^2 + \theta_{23}^2 +
\theta_{31}^2) M^2} \right ) \right ] , \hspace{3cm} \eqno{(3.11)}$$
\\
Therefore, when the thermal wavelength is smaller than the noncommutativity scale,
i.e.   

$${\beta^2 \over (\theta_{12}^2 + \theta_{23}^2 + \theta_{31}^2) M^2} <1,$$ 
\\
then the first term in Eq.(3.11) is a good approximation.    In a similar way we
have the following approximation

$$(3.10) \approx {1\over\beta^2}\int _0^1 dw \int_0^\infty d\alpha_1 \int_0^\infty
d\alpha_2  {\pi \sqrt\alpha_1\over
\sqrt{\alpha_2}} ~ {4 \pi^2\over (\theta_{12}^2 + \theta_{23}^2 + \theta_{31}^2)}
~e^{-(\alpha_1+\alpha_2) M^2} \times \hspace{5cm}$$
$$  \hspace{4cm}{\displaystyle \sum_{k_0 }}{\displaystyle \sum_{p_0 }} ~
e^{-[\alpha_2 + \alpha_1(w-w^2)]({2\pi p_0\over \beta })^2} ~ e^{-\alpha_1[{2\pi
k_0\over \beta }+(1-w){2\pi p_0\over \beta }]^2} \hspace{2cm}$$

$$ = {2 \pi^4\over  \theta_{12}^2 + \theta_{23}^2 + \theta_{31}^2} ~ {1\over M^4
\beta^2} . \hspace{3cm}\eqno{(3.12)}$$
\\
Substituting the above results into Eqs.(3.3) and (3.4) we finally find
the nonplanar-diagram contributions of the effective potential

$$ V(\phi_0)^{nonplanar}_{two~loop}  \approx { \lambda \hbar ^2 \over 1152 \pi^2} ~
{1  \over {\theta_{12}}^2 + \theta_{23}^2 + \theta_{31}^2 }~ {1\over\beta^2} ~  ({
1\over M^2} - {\lambda  \phi_0^2 \over 2 M^4})  \hspace{3cm} \eqno{(3.13).}$$

   
\subsection {Planar Diagrams}

   The planar-diagram contributions of Eqs.(3.1) and (3.2) can be analyzed
following Eqs.(3.9) and (3.10) after letting $\theta_{ij} =0$.   Thus, Eq.(3.1) with
$\theta_{ij} =0$ becomes

$$ {\lambda \over \beta^2}~ \int_0^\infty d\alpha_1 \int_0^\infty d \alpha_2 ~
{\pi^3\over ( \alpha_1 \alpha_2)^{3\over2}} ~e^{-(\alpha_1+\alpha_2) M^2} ~
{\displaystyle \sum_{p_0 }} e^{-\alpha_1 ({2\pi p_0\over\beta})^2}  ~{\displaystyle
\sum_{l_0 }} e^{-\alpha_1 ({2\pi l_0\over\beta })^2} $$

$$ = {\lambda \over \beta^2} ~  \pi^3 \left[\int_0^\infty d\alpha ~
\alpha^{\epsilon} \alpha^{-{3\over2}} ~e^{-\alpha M^2} ~ {\displaystyle \sum_{p_0 }}
e^{-\alpha ({2\pi p_0\over\beta})^2} \right]^2 $$

$$= {\lambda \over \beta^2} ~ \pi^3 \left[ \Gamma (-{1\over2}+\epsilon) ~
{\displaystyle \sum_{p_0 }} ~ [({2\pi p_0\over\beta})^2 + M^2]^{{1\over2}-\epsilon}
\right]^2 $$

$$ = {\lambda \over \beta^2} ~ \pi^3 ~ [\Gamma (-{1\over2})]^2 \left[
\zeta({-1}){4\pi M\over\beta} + M^2 \left (1 + O(\beta^2 M^2 \right ) \right],
\eqno{(3.14)}$$
\\
which is consistent with that evaluated by Dolan and Jackiw [20].   Note that in
above (and in the following, if it is necessary) we have added $\alpha ^{\epsilon}$
to regularize the integration over $\alpha$.   The infinite term appearing in the
above equation has also been neglected, as it will give a temperature-independent
contribution to the effective potential and could be absorbed by the counterterms
for the renormalization requirement [20]. 

    Note that, as it is difficult to evaluate Eq.(3.2) the authors in [20]
considered the $N$-component scalar field theory, i.e. $\phi \to \phi_a$ with $a =
1,2 ...N$.  In this system the contributions of diagram (3.1) is O($N^2$) and
diagram (3.2) O($N$).   Thus in the large $N$ approximation the diagram (3.2) can be
neglected and the two-loop correction calculated in [20] is consistent with our
result in Eq.(3.14).   However, in obtaining the above result, as we have used the
zeta regularization method [23] to perform the summation over the integral values of
 $p_0 $ the calculations become easy and, in a similar way, we could also evaluate
Eq.(3.2) in the high-temperature approximation.

   Using the same method, Eq.(3.2) with $\theta_{ij} =0$becomes

$$ {\lambda^2 \phi^2_0 \over \beta^2} ~ \int _0^1 dw \int_0^\infty d\alpha_1
\int_0^\infty d\alpha_2 {\pi \sqrt\alpha_1\over \sqrt{\alpha_2 +\alpha_1(w-w^2)}} ~
{\pi^2\over  \alpha_1 (\alpha_2+\alpha_1(w-w^2)) } ~e^{-(\alpha_1+\alpha_2) M^2}
\times\hspace{1cm}$$
$${\lambda^2 \phi^2_0 \over \beta^2}~   {\displaystyle \sum_{k_0 }}{\displaystyle
\sum_{p_0 }} ~ e^{-[\alpha_2 + \alpha_1(w-w^2)]({2\pi p_0\over \beta })^2} ~
e^{-\alpha_1[{2\pi k_0\over \beta }+(1-w){2\pi p_0\over \beta }]^2} $$
$$\approx {\lambda^2 \phi^2_0 \over \beta^2}~  \int _0^1 dw \int_0^\infty d\alpha_1
\int_0^\infty d\alpha_2 {\pi \sqrt\alpha_1\over \sqrt{\alpha_2 }} ~ {\pi^2\over 
\alpha_1 \alpha_2} ~e^{-(\alpha_1+\alpha_2) M^2} {\displaystyle \sum_{k_0
}}{\displaystyle \sum_{p_0 }} ~ e^{- \alpha_2 ({2\pi p_0\over \beta })^2} ~
e^{-\alpha_1[{2\pi k_0\over \beta }+(1-w){2\pi p_0\over \beta }]^2} $$
$$= {\lambda^2 \phi^2_0 \over \beta^2}~ \pi ^3 \Gamma(-{1\over2})  ~
\Gamma({1\over2})\int _0^1 dw ~  {\displaystyle \sum_{k_0 }}{\displaystyle \sum_{p_0
}} ~\left[({2\pi p_0\over \beta })^2  + M^2\right]^{-{1\over2}} \left[[{2\pi
k_0\over \beta }+(1-w){2\pi p_0\over \beta }]^2 + M^2\right]^{1\over2} $$
$$ = {\lambda^2 \phi^2_0 \over \beta^2}~ {\pi ^3 \over 4}\Gamma(-{1\over2})
~\Gamma({1\over2})~\left[1 + O(\beta^2 M^2) \right] . \hspace{8.5cm}\eqno{(3.15)}$$

   Substituting the above results into Eqs.(3.1) and (3.2) we finally find
the planar-diagram contributions of the effective potential

  $$V(\phi_0,N )^{planar}_{two~loop} = - \lambda \hbar^2 {M \over 293 \pi  \beta^3 }
~ \left (1 + O(\beta^2 M^2 )\right ).  \eqno{(3.16)}$$


\subsection{Effective Potential and Symmetry Property}

Combining Eqs.(3.13) and (3.16),  we find that the high-temperature effective
potential is 

$$ V(\phi_0)_{two ~ loop}  \approx  - \lambda \hbar^2 {M \over 293\pi \beta^3 }+ {
\lambda \hbar ^2 \over 1152 \pi^2} ~ {1  \over ({\theta_{12}}^2 + \theta_{23}^2 +
\theta_{31}^2)~\beta^2}~ ({ 1\over M^2} - {\lambda  \phi_0^2 \over 2 M^4}) . 
\eqno{(3.17)}$$
\\
From the above result we see that at high temperature in which the thermal
wavelength is smaller than the noncommutativity scale the planar-diagram
contributions will dominate.  This is the conclusion of this paper. The remaining
part of this paper is to investigate the property of the nonplanar contribution.    
From Eq.(3.13) we see that 

$$ {\partial V(\phi_0)^{nonplanar}_{two~loop}\over \partial (\phi_0^2)} \approx  {
\lambda \hbar ^2 \over 1152 \pi^2} ~ {- \lambda m^2 \over ({\theta_{12}}^2 +
\theta_{23}^2 + \theta_{31}^2)~\beta^2 M^6} . \eqno{(3.18)}$$ 
\\
The above equation tells us that, if  $ m^2 > 0 $, i.e., the symmetry is not broken
in the tree level, then  the value $\partial V(\phi_0)^{nonplanar}_{two~loop} /
\partial (\phi_0^2)$  becomes negative for all values of $\phi_0^2$.    This means
that the nonplanar diagram has an inclination to induce radiatively symmetry
breaking  if it is not broken in the tree level.  

  On the other hand,  if  $m^2 < 0$, i.e., the symmetry has been spontaneously
broken in the tree level, then because the value of $M^6 =( m^2 + {1\over2}\lambda
\phi_0^2)^3$ can become negative for small value of $\phi^2_0$ we therefore see that
the value $\partial V(\phi_0)^ {nonplanar}_{two~loop} / \partial (\phi_0^2)$  in
Eq.(3.18) is  negative too.  This means that the nonplanar diagram does bot have an
inclination to restore the symmetry breaking at high temperature, in contrast to the
conventional believing that high temperature could restore the symmetry breaking.  
Note that the investigation in [20] had found that the O(N)-invariant scalar model
which is radiatively broken will remain broken at high temperature, while which is
broken at tree level can be restored at high temperature.    

    In the next subsection we will evaluate the planar diagrams in zero space.  We
will see that the nonplanar-diagram corrections in the three space behave like those
in the planar diagrams in zero space, at high temperature. Thus we can explain this
property as a consequence of the drastic reduction of the degrees of freedom in the
nonplanar diagrams at high temperature.


\subsection {Planar Diagram at Zero space and Dimensional Reduction of Nonplanar
Diagram}

    The calculations of diagrams (3.1) and (3.2) at zero space dimension are 

$$[Diagram ~ (3.1) ]_{zero~space~dimension} =  ~{2\over 3}~{{\hbar ^2} \over 24}
{\lambda \over\beta^2}~  \left ({\displaystyle \sum_{k_0 }} {1\over({2\pi
k_0\over\beta })^2 + M^2}\right )^2 $$
$$=  ~{2\over 3}~{{\hbar ^2} \over 24} {\lambda \over\beta^2 M^4}~  \left( 1 +
O(\beta^2 M^2)~\right ). \eqno{(3.19)}$$

$$ [Diagram ~ (3.2) ]_{zero~space~dimension}= - ~ {1\over 2} ~ {{\hbar ^2} \over 36}
{\lambda ^2\phi_0^2 \over \beta^2}{\displaystyle \sum_{k_0 }}{\displaystyle
\sum_{p_0 }}{1\over({2\pi k_0\over \beta})^2 + M^2} \times$$
$$  {1\over({2\pi p_0\over \beta})^2 + M^2} {1\over({2\pi k_0\over \beta}+ {2\pi
p_0\over \beta})^2 + M^2}$$
$$ = - ~ {1\over 2} ~ {{\hbar ^2} \over 36}  {\lambda ^2\phi_0^2 \over \beta^2 M^6}
\left( 1 + O(\beta^2 M^2) \right),  \eqno{(3.20)}$$
\\
in which the zeta regularization method has been used to perform the summations over
integral values $p_0$ and $k_0$.  (Note that there is no nonplanar diagram in the
zero-space theory.)

   The important difference between the theory at  three spaces dimension and that
at zero space is that, in the former system the contribution of the planar-diagram
part of Eq.(3.1), calculated in Eq.(3.14), is proportional to $\beta ^{-3}$ while
that in the later system, calculated in Eq.(3.19), is proportional to $ \beta
^{-2}$.   The calculation in Eq.(3.11), however, find  that the high-temperature
two-loop correction from the nonplanar diagrams in three-space theory is
proportional to $ \beta ^{-2}$, which therefore behaves as the Feynman diagram in
zero-space theory.  

    In the same way of comparison, the high-temperature two-loop correction from the
nonplanar diagram of Eq.(3.4) in three-space theory, calculated in Eq.(3.12),  is
proportional to $ \beta ^{-2}$, which  does behave like as the Feynman diagram in
zero-space theory, calculated in Eq.(3.20).  

   Therefore, we can conclude that there is the drastic reduction of the degrees of
freedom in the nonplanar diagram at high temperature. This is because that, at high
temperature for which the thermal wavelength is small than the noncommutativity
scale, there is no way to distinguish and count the contributions of modes to the
nonplanar-diagram contribution in the effective potential.   To confirm the above
argument we would like to analyze the  $\lambda \phi^3$ theory in the next section.


\section {Two-Loop Corrections: $\lambda \phi^3$ Theory}

    It is known that in $\lambda \phi^3$ theory there is only one diagram in  the
two-loop level, as that given in Eq.(3.2).   Thus we need only to analyze this
equation in this section.   For the $\lambda \phi^3$ theory in three space
dimensions it is easy to see that, up to an irrelevant constant, the contributions
from the planar diagram and the nonplanar diagram  are like that in the $\lambda
\phi^4$ theory in three space dimensions.     Thus the property of the reduction of
degrees of freedom at high temperature is also shown in this system.  

   Let us now turn to the $\lambda \phi^3$ theory in five space dimensions.   The
space noncommutative, for simplicity, is chosen as 

$$ \left(\begin {array}{cccccc}
0&0&0&0&0&0\\
0&0&\theta_{12}&\theta_{13}&0&0\\
0&\theta_{21}&0&\theta_{23}&0&0\\
0&\theta_{31}&\theta_{32}&0&0&0\\
0&0&0&0&0&\theta\\
0&0&0&0&- \theta&0\\
\end{array}\right),  \eqno{(4.1)}$$
\\
which means that the extra noncommutative two spaces, with noncommutativity 
$\theta$, are commutate to the other noncommutative three spaces.

   Following the prescriptions in section 3 we can find from Eq.(3.6) the
coresponding contribution of the nonplanar diagram:  

$$ {\lambda ^2\phi_0^2 \over \beta^2} ~ {\displaystyle \sum_{k_0 }}{\displaystyle
\sum_{p_0 }}\int d^5 {\bf k} d^5 {\bf p} {1\over({2\pi k_0\over \beta})^2 + {\bf
k}^2 +M^2} {1\over({2\pi p_0\over \beta})^2 + {\bf p}^2 +M^2} {e^{i {\bf
k}_i\theta^{ij} {\bf p}_j}\over({2\pi k_0\over \beta}+ {2\pi p_0\over \beta})^2 +
({\bf p +\bf k})^2 +M^2} $$
$$\approx {\lambda ^2\phi_0^2 \over \beta^2} ~ \int _0^1 dw \int_0^\infty d\alpha_1
\int_0^\infty d\alpha_2 {\pi \over \sqrt{\alpha_1}} ~ {\pi\over \sqrt{\alpha_2}}~{4
\pi^2\over (\theta_{12}^2 + \theta_{23}^2 + \theta_{31}^2)} ~ {\pi^2\over \theta^2}
~ e^{-(\alpha_1+\alpha_2) M^2} \times \hspace{3cm}$$
$$  \hspace{4cm}{\displaystyle \sum_{k_0 }}{\displaystyle \sum_{p_0 }} ~
e^{-[\alpha_2 + \alpha_1(w-w^2)]({2\pi p_0\over \beta })^2} ~ e^{-\alpha_1[{2\pi
k_0\over \beta }+(1-w){2\pi p_0\over \beta }]^2} \hspace{2cm}$$
$$ = {\lambda ^2\phi_0^2 \over \beta^2}  ~ {4 \pi^6\over  \theta_{12}^2 +
\theta_{23}^2 + \theta_{31}^2} ~ {1\over \theta^2 M^2 }  \left[ 1 + O\left({\beta^2
\over (\theta_{12}^2 + \theta_{23}^2 + \theta_{31}^2) M^2} \right ) \right ].
\hspace{3cm}\eqno{(4.2)}$$
\\
   In a similar way,  we can find from Eq.(3.15) the coresponding contribution of
the planar diagram:

$$ {\lambda ^2\phi_0^2 \over \beta^2} ~ {\displaystyle \sum_{k_0 }}{\displaystyle
\sum_{p_0 }}\int d^5 {\bf k} d^5 {\bf p} {1\over({2\pi k_0\over \beta})^2 + {\bf
k}^2 +M^2} {1\over({2\pi p_0\over \beta})^2 + {\bf p}^2 +M^2} {1\over({2\pi k_0\over
\beta}+ {2\pi p_0\over \beta})^2 + ({\bf p +\bf k})^2 +M^2} $$
$$ \approx {\lambda ^2\phi_0^2 \over \beta^2} ~ \int _0^1 dw \int_0^\infty d\alpha_1
\int_0^\infty d\alpha_2 {\pi^5 e^{-(\alpha_1+\alpha_2) M^2}\over
({\alpha_1\alpha_2})^{3\over2}} ~ {\displaystyle \sum_{k_0 }}{\displaystyle
\sum_{p_0 }} ~ e^{-[\alpha_2 + \alpha_1(w-w^2)]({2\pi p_0\over \beta })^2} ~
e^{-\alpha_1[{2\pi k_0\over \beta }+(1-w){2\pi p_0\over \beta }]^2} $$
$$ = {\lambda ^2\phi_0^2 \pi^6\over \beta^2} ~~ \left [ -  {2\pi M \over \beta} +
M^2 \left (1+ O(\beta^2 M^2) \right ) \right ].  \hspace{7cm} \eqno{(4.3)}$$
\\
We thus see that the high temperature  behavior of the nonplanar diagram for the
$\lambda \phi^3$ theory living on five space dimensions, i.e. Eq.(4.2), has a
similar temperature dependence to that living on zero space dimension, i.e.
Eq.(3.20).   Note that all the properties found in section 3 could also be shown in
the $\lambda \phi^3$ at five space dimensions.
 

\section {Conclusion}

    In this paper we have evaluated  the high-temperature renormalized effective
potential for the scalar field theory in the noncommutative spacetime to the
two-loop approximation.  We have considered the $\lambda \phi^4$ theory in the three
space dimensions and $\lambda \phi^3$ theory in the five space dimensions.   As the
space-time noncommutativity ($\theta_{0i} \ne 0$) will lead to a non-unitary theory
[18], we have considered only space noncommutative theories. 

   Using the path-integration formulation we see that there is no nonplanar diagram
in the one-loop potential.   Thus the spontaneous symmetry breaking is blind to the
noncommutativity at this level.  The nonplanar parts can appear in the two-loop
level but they do not have an inclination to restore the symmetry breaking in the
tree level.  This property is in contrast to the conventional believing that high
temperature could restore the symmetry breaking.  To explain this phenomena we
compare the result with the effective potential in the zero-space dimension and
explain this property as a consequence of the drastic reduction of the degrees of
freedom in the nonplanar diagram at high temperature in which the thermal wavelength
is smaller than the noncommutativity scale [15].   Our results show that the
nonplanar two-loop contribution to the effective potential can be neglected in
comparsion with that from the planar diagrams.

   In this paper we only consider the noncommutative scalar field theory.   For the
more realistic model, such as that  including the Yang-Mills field, there are more
diagrams to be evaluated to find the effective potential.   However, we believe that
the property of the reduction of the degrees of freedom in the nonplanar diagram at
high temperature will also be shown in other models.   It remains to be
investigated.

\newpage
\begin{enumerate}

\item  A. Connes, M. R. Douglas and A. Schwarz,  {\it Noncommutative Geometry and
Matrix Theory: Compactification on Tori}, JHEP {\bf 02} (1998) 003,
hep-th/9711162;\\
M.R. Douglas, C. Hull, {\it D-branes and the Noncommutative Torus},
JHEP {\bf 02} (1998) 008,  hep-th/9711165.

\item M. M. Sheikh-Jabbari, {\it More on Mixed Boundary Conditions and D-branes
Bound States}, Phys. Lett.{\bf B 425} (1998) 48, hep-th/9712199;\\
F. Ardalan, H. Arfaei, M.M. Sheikh-Jabbari, {\it Mixed Branes and M(atrix)
Theory on Noncommutative Torus}, hep-th/9803067; {\it Noncommutative Geometry From
Strings and Branes}, JHEP {\bf 02} (1999) 016, hep-th/9810072;
{\it Dirac Quantization of Open Strings and Noncommutativity in Branes},
 Nucl. Phys. {\bf B576} (2000) 578, hep-th/9906161.
\item  N.~Seiberg and E.~Witten, {\it String Theory and Noncommutative Geometry}, 
JHEP {\bf 09}, 032 (1999), hep-th/9908142.
\item C-S. Chu, P-M. Ho, {\it Noncommutative Open String and D-brane}, Nucl. Phys.
{\bf B550} (1999) 151, hep-th/9812219; {\it Constrained Quantization of Open String
in background B Field and Noncommutative D-brane}, Nucl. Phys. {\bf B568} (2000)
447, hep-th/9906192;\\ 
T. Lee, {\it Canonical Quantization of Open String and Noncommutative Geometry},
Phys. Rev {\bf D 62} (2000) 024022,  hep-th/9911140.
\item 
T. Filk, {\it Divergences in a Field Theory on Quantum Space},   Phys.  Lett. {\bf
B376} (1996) 53;\\
B. A. Campbell and K. Kaminsky, {\it Noncommutative Field Theory and Spontaneous
symmetry breaking},  Nucl. Phys. {\bf B 581} (2000) 240, hep-th/0003137.
\item S. Minwalla, M. Van Raamsdonk, N. Seiberg, {\it Noncommutative Perturbative
Dynamics}, JHEP {\bf 02} (2000) 020,  hep-th/9912072.
\item I. Ya. Aref'eva, D.M. Belov, A.S. Koshelev, {\it Two-Loop
Diagrams in $\phi^4_4$ Theory},  Phys. Lett. {\bf B476} (2000) 431,
hep-th/9912075;\\
A. Micu and M. M. Sheikh-Jabbari, {\it Noncommutative $\phi^4$ Theory at Two loop},
hep-th/0008057.
\item W. H. Huang,  {\it Two-Loop Effective Potential in Noncommutative scalar field
theory},  Phys. Lett.{\bf B 496} (2000) 206, hep-th/0009067.
\item  C.P. Martin, D. Sanchez-Ruiz, {\it The One-loop UV Divergent Structure of
U(1) Yang-Mills Theory on Noncommutative $R^4$}, Phys. Rev. Lett. {\bf 83} (1999)
476, hep-th/9903077.
\item M.M. Sheikh-Jabbari, {\it One Loop Renormalizability of Supersymmetric
Yang-Mills Theories on Noncommutative Two-Torus}, JHEP {\bf 06} (1999) 015,
hep-th/9903107;\\
T. Krajewski, R. Wulkenhaar, {\it Perturbative quantum gauge fields on the
noncommutative torus}, Int. J. Mod. Phys. {\bf  A15} (2000) 1011, hep-th/9903187;\\
A. Matusis, L. Susskind, N. Toumbas, {\it The IR/UV Connection in the
Noncommutative Gauge Theories}, JHEP {\bf 12} (2000) 002 , hep-th/0002075;\\
M. Hayakawa, {\it Perturbative analysis on infrared aspects of noncommutative QED on
$R^4$}, Phys. Lett. {\bf B478} (2000) 394, hep-th/9912094; {\it Perturbative
analysis on infrared and ultraviolet aspects of noncommutative QED on $R^4$},
hep-th/9912167.
\item H. S. Snyder, Phys. Rev. {\bf 71} (1947) 38;  {\bf 72} (1947) 68.
\item   Connes, {\it Noncommutative Geometry} Academic. Press, New York,
1994);
Connes A.{\it Gravity coupled with matter and the foundation of non
commutative geometry}, Comm. in Math. Phys. {\bf 182} (1996) 155, hep-th/9603053;\\ 
Landi G. {\it An introduction to noncommutative spaces and their
geometries}, Lecture Notes in Physics, Springer-Verlag , hep-th/97801078;\\
Sch\"ucker T. {\it Geometries and forces}, Summer School {\it Noncommutative
geometry and applications}, Lisbon September 1997,
hep-th/9712095 .
\item S. Nam, {\it Casimir Force in Compact Noncommutative Extra Dimensions
and Radius Stabilization}, JHEP {\bf 10} (2000) 044,
hep-th/0008083;\\
J. Gomis, T. Mehen and M.B. Wise, {\it Quantum Field Theories with Compact
Noncommutative Extra Dimensions}, JHEP {\bf 08} (2000) 029; hep-th/0006160;\\ 
W. H. Huang, {\it Casimir Effect on the Radius Stabilization of the 
Noncommutative Torus},   Phys. Lett.{\bf B 497} (2001) 317; hep-th/0010160;\\
W. H. Huang, {\it Finite-Temperature Casimir Effect on Radius Stabilization
of the  Noncommutative Torus}, JHEP {\bf 11} (2000) 041,
hep-th/0011037.
\item G. Arcioni and M. A. Vazquez-Mozo,  {\it Thermal effects in perturbative
noncommutative gauge theories}, JHEP {\bf 01} (2000) 028;\\
G. Arcioni, J.L.F. Barbon, Joaquim Gomis and M. A. Vazquez-Mozo,  {\it On the
stringy nature of winding modes in the noncommutative thermal field theories}, JHEP
{\bf 06} (2000) 038, hep-th/0004080.
\item W. Fischler, E. Gorbatov, A. Kashani-Poor, S. Paban, P. Pouliot and J. Gomis,
{\it Evidence for Winding States in Noncommutative Quantum Field Theory}, JHEP {\bf
05} (2000) 024, hep-th/0002067;\\
W. Fischler, E. Gorbatov, A. Kashani-Poor, R. McNees, S. Paban, P. Pouliot and J.
Gomis, {\it The interplay between $\theta$ and T}, JHEP {\bf 06} (2000) 032,
hep-th/0003216.
\item R.-G. Cai and N. Ohta, {\it On the thermodynamics of large N Noncommutative 
Super Yang-Mills Theory}, Phys. Rev. {\bf D 61} (2000) 124012, hep-th/9910092.
\item J. Gomis, K. Landateiner, and E. Lopez, {\it Nonrelativistic Non-commutative
Field Theories and UV/IR Mixing}, Phys. Rev. {\bf D 62} (2000) 105006,
hep-th/0004115. 
\item  J. Gomis and T. Mehen, {\it Spacetime  Noncommutative Field Theories and
Unitarity}, Nucl. Phys. {\bf B 591} (2000) 265, hep-th/0005129; \\
N. Seiberg, L. Susskind, N. Toumbas, {\it Space/Time Non-Commutativity and
Causality},  JHEP {\bf 06} (2000) 044, hep-th/0005015. 
\item R. Jackiw, Phys. Rev. {\bf D 9} (1974) 1686.
\item L. Dolan and R. Jackiw, Phys. Rev. {\bf D 9} (1974)3320.
\item H. Weyl. Z. Physik, {\bf 46} (1927) 1. 
\item J. E. Moyal, Proc. Cambridge Philo. Soc. {\bf 45} (1949) 99.
\item E. Elizalde, S. D. Odintsov, A. Romeo, A. A. Bytsenko, and Zerbini, {\it Zeta
Regularization Techniques with Applications},World Scientific, (1994).
\end{enumerate}
\end{document}